\documentstyle[aps,prl,epsfig,amssymb,balanced,times]{revtex}
\begin{document}
\title{Coherence Resonance in Chaotic Systems}
\author{Carlos Palenzuela$^1$, Ra\'ul Toral$^{1,2}$, Claudio R. Mirasso$^1$,
Oscar Calvo$^1$ and James D. Gunton$^{2,3}$}
\address{$^1$Departament de F{\'\i}sica, Universitat de les Illes Balears,
E-07071 Palma de Mallorca, Spain\\
$^2$Instituto Mediterr\'aneo de Estudios Avanzados (IMEDEA),
CSIC-UIB, E-07071 Palma de Mallorca, Spain\\ $^3$Department of
Physics, Lehigh University, Bethlehem, PA 18015\footnote{Permanent Address}\\
}
\maketitle

\begin{abstract}
We show that it is possible for chaotic systems to display the main features of
coherence resonance. In particular, we show that a Chua model, operating in a
chaotic regime and in the presence of noise, can exhibit oscillations whose
regularity is optimal for some intermediate value of the noise intensity. We
find that the power spectrum of the signal develops a peak at finite frequency
at intermediate values of the noise. These are all signatures of coherence
resonance. We also experimentally study a Chua circuit and corroborate the
above simulation results. Finally, we analyze a simple model composed of two
separate limit cycles which still exhibits coherence resonance, and show that
its  behavior is qualitatively similar to that of the chaotic Chua system.\vspace{1.0truecm}

\end{abstract}

\begin{twocolumns}

When a dynamical system is subjected to an external periodic
forcing, it is a standard result that synchronization between the
system and the forcing can occur under a large variety of
conditions. A resonance is defined as the presence of a maximum in the
response of the system as a function of some control parameter (for
instance, the frequency of the external signal). Although one
might naively believe that fluctuations, either in the forcing or
in the intrinsic dynamics, will worsen the quality of the
synchronization, it is nowadays well established that, in some
cases, the response of a nonlinear dynamical system to an
external forcing can be {\sl enhanced} by the presence of noise
(fluctuations). The prototypical and pioneering example is that
of {\sl stochastic resonance}\cite{BSV81,NN81} by which a
bistable system under the influence of a periodic forcing,
and in the presence of fluctuations, shows an optimum
response, a resonance, for a given value of the noise intensity. The relevance
of this phenomenon has been shown for some physical and biological systems
described by a nonlinear dynamics\cite{JSP70,GHJM98,LC98}.

That noise can have a {\sl constructive} role has been one of the most
astonishing discoveries of the last decades in the field of stochastic
processes. Besides the above mentioned stochastic resonance, purely
temporal dynamical systems can display phenomena such as noise-induced
transitions\cite{HL84} or noise-induced transport\cite{HB96}. In spatially
extended systems, on the other hand, noise is known to induce a large
variety or ordering effects\cite{OS99}. In all the cases, the common
feature is that some sort of {\sl order} appears only in the presence of
the right amount of noise.

The possibility of having stochastic resonance without the need of an
external forcing has attracted much attention
recently\cite{GDNH93,RS94,PK97}. In particular, the phenomenon named {\sl
coherence resonance}\cite{PK97} was shown to appear in excitable systems
under the influence of fluctuations. An excitable system has a stable
fixed point with a finite basin of attraction. When a perturbation is such
that the system crosses a threshold value, the return to the fixed point
is by executing a large excursion in the configuration space, thus
generating a pulse in the time evolution. One of the main features of
excitable systems is that the generated pulse is basically independent of
the magnitude of the perturbation that induced its firing. Therefore, the
duration of the pulse, the excursion time $t_e$, is a characteristic of
the system and not of the perturbation. The total time between pulses,
$t_p$, is composed of two times: the excursion time $t_e$ and the time
needed for the activation of the pulse, $t_a$. If the firing of the pulses
is produced by random fluctuations, the activation time $t_a$ is a random
variable. According to Kramers formula, for small noise, the mean
activation time behaves as $\langle t_a \rangle \sim \exp(A/D^2)$ where
$A$ is a constant and $D$ is the noise intensity\cite{summ}. The variance
of the activation time is $\sigma^2[t_a]\approx \langle t_a \rangle^2$. At
the same time, the excursion time $t_e$ depends weakly on $D$, such that
its mean value $\langle t_e \rangle$ can be considered constant and its
variance can be estimated as $\sigma^2[t_e]\approx D^2 \langle t_e
\rangle$. For small $D$, we have $\langle t_a \rangle \gg \langle t_e
\rangle$ and we can approximate the time between pulses by the activation
time $t_p \approx t_a$. The relative fluctuations of the time between
pulses, defined as $R= \sigma[t_p]/\langle t_p \rangle$, is in this limit
of small noise $R \approx \sigma[t_a]/\langle t_a \rangle \approx 1$. For
large noise, the activation time is very small and the system fires a
pulse every time it returns from an excursion. The pulse time is dominated
by the excursion time and we can approximate $R \approx
\sigma[t_e]/\langle t_e \rangle \sim D  \langle t_e \rangle^{-1/2}$. If
the excursion time is large, and the threshold of excitation is small, it
is possible that for intermediate values of $D$, it is $R(D) < 1$. In this
case, and according to the generic behaviors described above ($R(D)\to 1$
for small $D$ and $R(D) \sim D\langle t_e \rangle^{-1/2}$ for large $D$)
there will be a minimum in the relative fluctuations of the time between
pulses. This is the signature of coherence resonance. A similar effect is that of stochastic
resonance without external periodic force which can occur in a system near
a limit cycle bifurcation point\cite{GDNH93,RS94}.

To summarize, the main feature of a system displaying coherence resonance
is that a quasi--periodic signal is generated by a combination of the
internal nonlinear dynamics and fluctuations without the need for the
presence of an external, deterministic, periodic signal. The periodicity
of the pattern is optimal (resonance) for a certain value of the noise
intensity. The original studies have been extended to consider other
excitable systems such as the FitzHugh-Nagumo model\cite{LS99,MV99}, the
Hodgkin-Huxley model for neurons\cite{LNK98} and the Yamada model for a
self-pulsating semiconductor laser\cite{DKL99}. Coherence resonance has
been also observed in dynamical systems close to the onset of a
bifurcation\cite{NSS97} as well as in other bistable and oscillatory
systems\cite{LS00,POC99}. Experimental evidence for the existence of
coherence resonance has been given for a laser system\cite{GGBT00} and for
excitable electronic circuits\cite{PHYS99,HYPS99}.

In this paper we will prove that it is possible to display the main
features of coherence resonance in chaotic and other bistable systems in
which the attractors are not of the fixed point type. Although we believe that our results
are quite general, we will consider specifically a Chua circuit operating
in a chaotic regime with two independent, symmetric, attractors. We will
argue that the existence of a well defined characteristic time when moving
around each attractor is a necessary ingredient for the ocurrence of
coherence resonance. This characteristic time plays the role of the
excursion time for excitable systems in the sense that only during a small
fraction of this time, when the trajectory comes as close as possible to the other attractor, the fluctuations can induce jumps between the two
attractors. We will give numerical and experimental evidence that such a Chua circuit in the presence of noise, can undergo oscillations whose
regularity is optimal (in a sense to be precisely defined later) for some
intermediate value of the noise intensity. Later, we will also show that
the basic ingredients for this new kind of coherence resonance are already
present in a simpler toy model with two separated limit cycles. That the addition of noise can induce some degree of regularity in a chaotic system has been shown recently in a different context related to the existence of noise--induced sychronization of chaotic systems\cite{TMHP00}

Let us consider the Chua system, in its dimensionless form, under the
presence of additive noise\cite{chua}:
\begin{eqnarray}
\dot x & = & \alpha(y-h(x))\nonumber\\
\dot y & = & x-y+z \label{eq1} \\
\dot z & = & -\beta y-\gamma z +\xi(t)\nonumber
\end{eqnarray}
where $\xi(t)$ is Gaussian white noise, of zero mean and correlations
$\langle \xi(t)\xi(t') \rangle = D^2 \delta(t-t')$. The nonlinear function
$h(x)$ is given by $h(x)=b x+\frac{a-b}{2}(|x+1|-|x-1|)$. We have taken
the values $a=-1/7$, $b=2/7$, $\alpha=4.60$, $\beta=6.02$, $\gamma=0$, for
which the Chua system has two chaotic attractors: a single scroll and its mirror image. Depending on the initial conditions, the system will rotate around one attractor or the other. In other words, in the absence of fluctuations, the attractors are independent and trajectories can not jump from one to the other. The movement around each attractor has a well defined mean angular frequency $\omega_0$, which for these values of the parameters is $\omega_0 \approx 3$.

\begin{figure}
\hspace{-1.0cm}\makebox{\epsfig{file=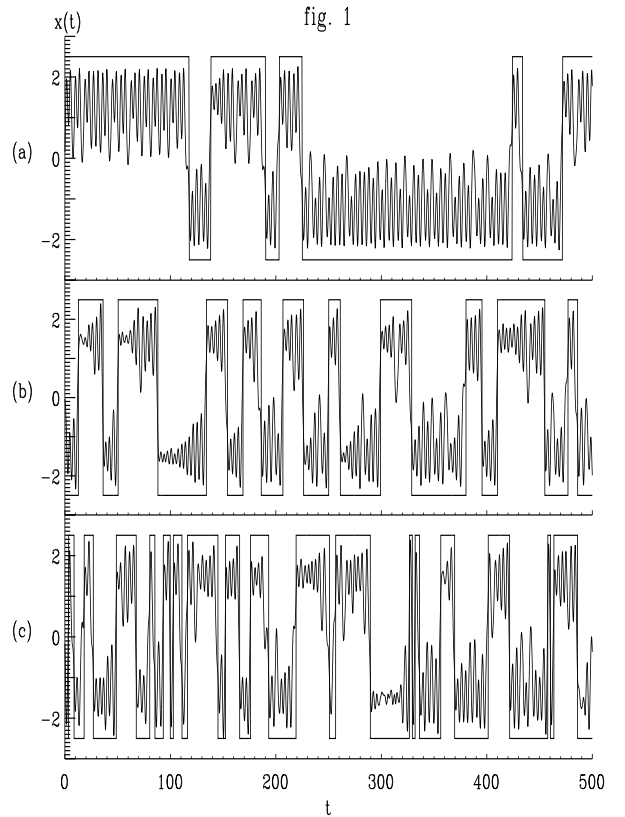,angle=90,
width=10.0cm, height=12.0cm}} \caption{Time series for the $x$ variable of
the Chua system given by Eqs. (\ref{eq1}) for three different noise levels:
a) $D=0.02$, b) $D=0.08$, optimum noise level, and c) $D=0.16$.
\label{fig1}} \end{figure}

\begin{figure}
\hspace{-1.0cm}\epsfig{file=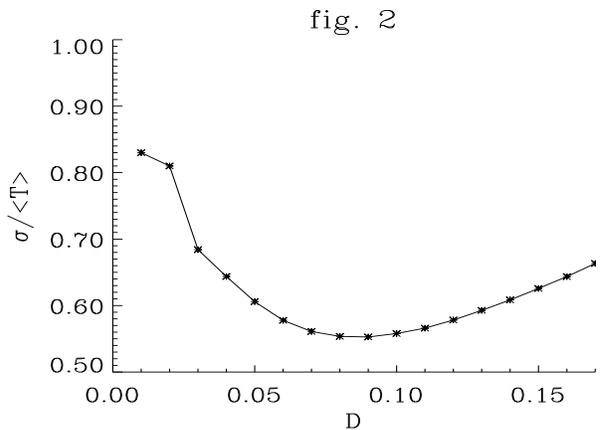,angle=90,
width=9.0cm, height=6.0cm}
\caption{Standard deviation normalized by
the mean time $\sigma/\langle T \rangle$ for the Chua system given by Eqs. (\ref{eq1}) for noise levels ranging
from 0.01 to 0.17.\label{fig2}}
\end{figure}

In figure \ref{fig1} we plot three trajectories of the variable $x(t)$
corresponding to increasing levels of the noise intensity $D$.  We observe
three qualitatively different behaviors when increasing the noise level.
When $D$ is very small (figure \ref{fig1}a) the average residence time,
i.e. the time between jumps, is large and the system spends most of the
time rotating around one of the attractors. For this situation, the
dispersion of the residence time is also large. As $D$ increases,
approaching an optimum value (figure \ref{fig1}b), the system jumps
between the two attractors more regularly. These jumps occur, as already
mentioned, approximately when the trajectory passes closest to the other
attractor. Finally, when $D$ is very large, the system jumps more often
but these jumps may start from different points on the trajectory and the
behavior of the system is more irregular (figure \ref{fig1}c). The
enhanced regularity that occurs for intermediate values of the noise can
be clearly observed in figure \ref{fig2} where we plot the standard deviation, $\sigma[T]$, of the residence time in each attractor, normalized
to its mean value, $\langle T \rangle$, as a function of the noise
intensity. This curve exhibits a minimum at a noise level $D \approx
0.08$. The presence of this minimum is the clearest signature of coherence
resonance. Another observed indicator of coherence resonance is the
existence of a peak in the the power spectrum $S(f)$ of the signal at
a finite frequency\cite{PK97,LS00}. Moreover, our data show a  maximum in
the ratio between the height and the width of the peak of $S(f)$ at
$D\sim 0.07$. Similarly, we observe that the time-correlation function, $C(t)$, has the longest tail and the lowest minimum at values of the noise close to the optimal level.

Although there is a good correspondence between a real Chua circuit and the system of differential
equations (\ref{eq1}), it is obvious that the numerical results deal necessarily with
an idealized Chua circuit. In order to analyze the robustness of the observed
phenomenon, we have performed experiments in a real Chua circuit constructed
according the classical design (see e.g. ref \cite{chua}) with the following
parameters\cite{exp}: $C_1=20$ nF, $C_2=100$ nF, $R=1100$ $\Omega$, $a=-1/7$,
$b=2/7$. The noise has been generated with a standard Hewlett-Packard function
generator and its intensity has been varied from zero to a few volts. As in the
numerical study, the original time series for the $x(t)$ variable has been converted
into a variable $u(t)$ taking the values $+1$ and $-1$ for each of the attractors.
Using this variable we have computed the normalized standard deviation of the
residence time, $\sigma[T]/\langle T \rangle$. In figure \ref{fig3} we plot this
quantity as a function of the noise intensity. Again, a clear minimum appears for an
optimal value $D\approx 5500$ mV, so confirming the numerical results. In figure
\ref{fig4} we plot the power spectrum of the digital variable $u(t)$. We notice the
development of a peak at a finite frequency for an intermediate noise level giving
further evidence of a regular behavior.

To gain more insight into the dynamics of this chaotic system we consider a
simplified two variable system $(x_1,x_2)$ with two stable limit cycles. The first
one, $C_1$, is around the unstable fixed point $(1,0)$ and the second one, $C_2$,
around the unstable fixed point $(-1,0)$. There are no other stable fixed points or
limit cycles in the system. We assume that the limit cycles are circumferences of
radius $R$ close to but smaller than $1$, and that they have a constant angular
speed $\omega_0$. Under these circumstances, which limit cycle is chosen as a
dynamical attractor depends exclusively on the initial condition\cite{equations}. Let
us add now some noise to the dynamics. If the noise intensity is small, the
modification to the trajectories will be small. Moreover, the probability that noise
induces a jump between the attractors is only significant near the closest points in
the limit cycles, i.e., when the trajectory passes closest to the origin of the
coordinate system. If the system does not jump at this point then it has to wait for
a complete rotation for another chance to jump. Hence, the rotation period $2
\pi/\omega_0$ plays the role of the excursion time $t_e$ in the excitable system in
the sense that, for moderate levels of noise, the system cannot jump to the other
attractor during this time.

\begin{figure}
\hspace{-1.0cm}\makebox{\epsfig{file=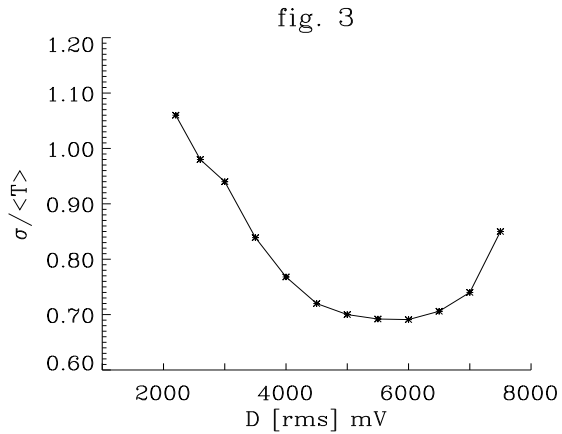,angle=90,
width=8.0cm, height=6.0cm}} \caption{Standard deviation normalized by
the mean time $\sigma/\langle T \rangle$ in the case of the Chua circuit (see the text for details of the parameters) for noise levels ranging
from $2000$ to $7500$ mV.\label{fig3}}
\end{figure}

\begin{figure}
\hspace{-1.0cm}\makebox{\epsfig{file=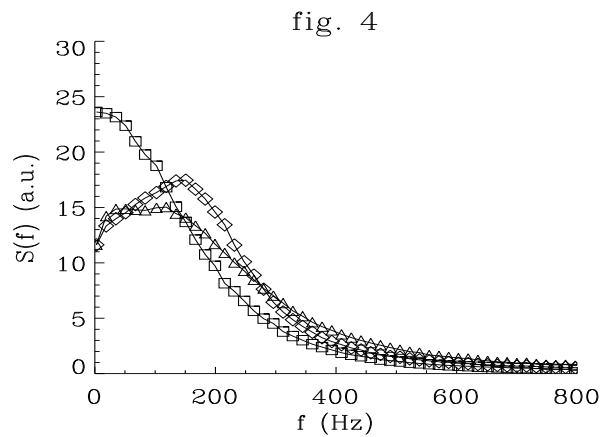,angle=90, width=9.0cm,height=6.0cm}} \caption{Power spectrum $S(f)$ of the digital signal $u(t)$ for the Chua circuit for different noise levels: $D=2000$ mV (squares), $D=4000$ mV (rhombi), $D=6000$ mV (triangles). The presence of a peak at a finite frequency is an evidence of coherence resonance.\label{fig4}}
\end{figure}

Let us define the variable $u={\rm sign}(x_1)$. This is equal to $+1$ when the system
is in attractor $C_1$ and $-1$ when in attractor $C_2$. The evolution of $u(t)$ can
be described by a series of time intervals alternating the values of $+1$ and $-1$.
According to the previous argument, for small noise, the duration $T$ of each time
interval is a random variable taking values which are an integer multiple of
$2\pi/\omega_0$. The probability that the system jumps between the attractors exactly
after $n$ cycles follows the geometric distribution: $P(T=2\pi
n/\omega_0)=p(1-p)^{n-1}$, where $p$ is the probability that the system jumps between
the two attractors during the time they are closest at each cycle. This probability
$p$ will be a function of the noise intensity $D$ and also of the angular speed
$\omega_0$ determining the time the system is ready for jumping. Using this
distribution, the relative fluctuation of the time between jumps is
$\sigma[T]/\langle T
\rangle = \sqrt{1-p}$. Therefore, for small $D$, $p$ will be small and $R$ will
initially decrease with $D$. Since, according to the general argument of \cite{PK97}
developed at the beginning, $R$ will eventually grow for large $D$, we expect a
minimum in a plot of $\sigma[T]/\langle T\rangle$ versus $D$. When reexamining
figures \ref{fig2} and \ref{fig3}, it can  also be observed that both limits, low and
high noise intensity, behave roughly as estimated for the simpler model described
above indicating that our simplified model captures the main ingredients of coherence
resonance in the more complicated chaotic system. However, there is an important
difference between the simple dynamical model and the Chua system. In the former, the
trajectory is always on the limit cycle and the system is ready to jump to the other
attractor at any cycle. On the contrary, in the Chua system, after the trajectory
jumps from one attractor to the other, the motion usually starts close to the center
of the attractor, in an inner orbit, and it is not ready to jump to the other
attractor until the outer orbits are reached. This fact can be guessed from figure
\ref{fig1} when looking at the times the system jumps from one attractor to the
other. This difference might be responsible from the fact that in the Chua model $R$
does not clearly approach to $1$ as $D$ is decreased. On the other hand, if noise is
arbitrarily increased, the Chua system saturates to a unique limit cycle thus losing
its characteristic behavior.

In conclusion, we have shown, both numerically and experimentally, that
coherence resonance can be observed in a chaotic system. We have also
shown that a simple model, composed of two separate limit cycles, is able
to exhibit coherence resonance. Within this model we were able to predict,
for instance, the limits of the normalized standard deviation of the residence
time by a simple analytical approximation.
The behavior of the chaotic Chua system follows qualitatively the results
derived in the simple model, with coherence resonance illustrated by the
dependence of several different quantities on the noise intensity.
Finally, we consider particularly interesting the fact that the
combination of noise and chaos can lead to some degree of regularity in the system.

{\bf Acknowledgements} This work has been supported by DGES
(Spain) projects PB94-0167, PB97-0141-C02-01 and NSF Grant DMR
9813409. JDG wishes to acknowledge the support of the Programa
C\'atedra of the Fundaci\'on BBVA and the Profesores Visitantes
of Iberdrola.

\end{twocolumns}


\begin{references}
\bibitem{BSV81} R. Benzi, A. Sutera and A. Vulpiani, J. Phys. {\bf A14}, 453 (1981).

\bibitem{NN81} C. Nicolis and G. Nicolis, Tellus {\bf 33}, 225 (1981).

\bibitem{JSP70} Proceedings of the NATO Advanced Research Workshop: Stochastic Resonance in Physics and Biology. F. Moss, A. Bulsara and M.F. Shlesinger, eds. J. Stat. Phys. {\bf 70} (1993).

\bibitem{GHJM98} L. Gammaitoni, P. H\"anggi, P. Jung and F. Marchesoni, Rev. Mod. Phys. {\bf 70}, 223 (1998).

\bibitem{LC98} A. Longtin and D.R. Chialvo, Phys. Rev. Lett. {\bf 81}, 4012 (1998).

\bibitem{HL84}  W. Horsthemke, R. Lefever, {\it Noise-Induced
Transitions} (Springer, Berlin, 1984).

\bibitem{HB96} P. H\"anggi and R. Bartussek, in {\sl Nonlinear Physics of Complex Systems}, edited by J. Parisi, S.C. M\"uller and W. Zimmermman (Springer, New York, 1999).

\bibitem{OS99} This field is growing too rapidly to list here the most
significant contributions. For a partial review, see
J.~Garc\'{\i}a--Ojalvo and J.M.~Sancho, {\em Noise in Spatially Extended
Systems} (Springer, New York, 1999).

\bibitem{GDNH93} H. Gang, T. Ditzinger, C.Z. Ning and H. Haken, Phys. Rev. Lett. {\bf 71}, 807 (1993).

\bibitem{RS94} W. Rappel and S. Strogatz, Phys. Rev. {\bf E50}, 3249 (1994).

\bibitem{PK97} A.S. Pikovsky and J. Kurths, Phys. Rev. Lett. {\bf 78}, 775 (1997).

\bibitem{summ} We are summarizing here the argument developed in \cite{PK97}

\bibitem{LS99} B. Lindner and L. Schimansky-Geier, Phys. Rev. E {\bf 60}, 7270 (1999).

\bibitem{MV99} S. Ripoll Massan\'es and C. J. P\'erez Vicente, Phys. Rev. E {\bf 59} 4490 (1999).

\bibitem{LNK98} S.G. Lee, A. Neiman and S. Kim, Phys. Rev. E {\bf 57}, 3292 (1998).

\bibitem{DKL99} J.L.A. Dubbeldam, B. Krauskof and D. Lenstra, Phys. Rev. {\bf E60}, 6580 (1999).

\bibitem{NSS97} A. Neiman, P. Saparin and L. Stone, Phys. Rev. E {\bf 56}, 270 (1997).

\bibitem{LS00} B. Lindner and L. Schimanwky-Geier, Phys. Rev. E {\bf 61}, 6103 (2000).

\bibitem{POC99} J.R. Pradines, G.V. Osipov and J.J. Collins, Phys. Rev. E {\bf 60}, 6407 (1999).

\bibitem{GGBT00} G. Giacomelli, M. Giudici, S. Balle and J.R. Tredicce, Phys. Rev. Lett. {\bf 84}, 3298 (2000).

\bibitem{PHYS99} D.E. Postnov, S.K. Han, T.G. Yim and O.V. Sosnovtseva, Phys. Rev. {\bf E59}, R3791 (1999).

\bibitem{HYPS99} S.K. Han, T.G. Yim, D.E. Postnov and O.V. Sosnovtseva, Phys. Rev. Lett. {\bf 83}, 1771 (1999).

\bibitem{TMHP00} R. Toral, C. Mirasso, E. Hern\'andez-Garc{\'\i}a and O. Piro, in {\sl Unsolved Problems on Noise and Flcutuations, UPoN'99}, D. Abbott and L. Kiss, eds. {\bf 511}, 255, American Insitute of Physics (Melville, NY, 2000).

\bibitem{chua} {\sl Chua's Circuit: A Paradigm for Chaos}, R.N. Madan, ed. World Scientific Publishing (1993).

\bibitem{exp} Full details of the experiments will be given elsewhere.

\bibitem{equations} A specific model with this kind of behavior is given by the dynamical system:
\begin{eqnarray*}
\dot r & = & -2r(r-R)\\
\dot \theta & = & \omega_0
\end{eqnarray*}
Where $r=\sqrt{(x_1-s)^2+x_2^2}$, $\theta=\arctan\left(\frac{x_2}{x_1-s}\right)$, and
$s=1$ for $x_1\ge 0$ and $s=-1$ for $x_1< 0$. A similar behavior occurs in a model for a Degenerate Optical Parametric Oscillator, where the phenomenon of coherence resonance can also be observed (I. Rabiosi, R. Toral and M. San Miguel, work in progress).

\end{references}
\end{document}